\documentstyle[aps,preprint]{revtex}
\tightenlines
\begin{document}
\draft

\title{Accuracy of a mechanical single electron shuttle}

\author{C.\ Weiss and  W.\ Zwerger} 
\address{Sektion Physik, Ludwig-Maximilians-Universit\"at M\"unchen, Theresienstr.~37,\\ D-80333 M\"unchen, Germany}
  
\date{\today}

\maketitle

\begin{abstract}
 Motivated by recent experiments,  we calculate both the average current and the current fluctuations for a metallic island which oscillates between two symmetric electrodes. Electrons can only tunnel on or off the island when it is close to one of the electrodes. Using a Master equation we investigate the accuracy of such an electron shuttle both analytically and numerically. It is shown that optimum operation is reached when the contact time is much larger than the $RC$-time. 
\end{abstract}

\section{Introduction}
One of the most exciting prospects of Coulomb blockade is the realization of a current standard based on periodically transferring single electrons through the device \cite{Gee90,Pot92}. In order to minimize higher order tunnelling processes which limit the accuracy, a chain of tunnelling junctions was used \cite{Kel96}.
Quite recently, a single electron shuttle mechanism which is based on mechanical oscillations between two electrodes was suggested by Gorelik {\it et al.\/} \cite{Gor98}.  The major advantage of a mechanically driven single electron transfer is that higher order processes are negligible.

In the original suggestion of a mechanical single electron shuttle a metallic island is connected via organic molecules to two electrodes. One of the electrodes is positive and the other is negatively biased. For mechanically soft molecules self-excitation is expected to internally drive oscillations of the island which transfers charge at frequency $f$. Because of Coulomb blockade an integer number of electrons is loaded onto the grain close to one turning point, and the same number of electrons is unloaded close to the other. The number of electrons transferred per cycle is thus expected to be
\begin{equation}
  \left<N\right>= 2n_{\rm max}\,, \hspace{1cm} n_{\rm max} =\left[\frac{CV}{e}+\frac 12\right]\,,
\label{eq:N_average}
\end{equation}
where $V$ is the applied voltage and $C$ the island capacitance relative to the electrodes. The average current is given by $ \left<I\right>= ef\left<N\right>\,$. If the island is situated between the two electrodes there should be either $n_{\rm max}$ or $-n_{\rm max}$ electrons on it depending on which electrode it is coming from.

A first step towards an experimental realization of a mechanical single electron shuttle was undertaken by Erbe {\it et al.\/}  \cite{Erb98}. The authors designed a semiconductor device where a mechanical clapper oscillates between two electrodes - {\it i.e.\/} the oscillation is now externally driven with the average number of electrons transferred in each cycle of order $7\pm 2$ \cite{Erb98}. This experiment which works similar to a bell is still in the classical regime, {\it i.e.\/} Coulomb blockade is not relevant here. The aim in a future experimental setup with a metallic island on the tip of the clapper is to obtain an electron shuttle mechanism which is limited to single electrons by Coulomb repulsion. 

Since the tunnelling resistance $R$ increases exponentially with the distance $\Delta x$ between the island and the electrodes ($R=\exp(\Delta x/\lambda)$), tunnelling is suppressed exponentially when the island is in the middle between the two electrodes (both the oscillation amplitude and the distance between the two electrodes is much larger than $\lambda$). For the same reason both cotunnelling and simultaneous tunnelling between the island and the electrodes can be neglected. In order for electrons to tunnel the amplitude of the oscillation has to be large compared with the typical tunnelling distance $\lambda$. In the experiments by Erbe {\it et al.\/} the oscillation amplitude is of order 100 nm (compared to a few nanometers as suggested by Gorelik {\it et al.\/}) and thus much larger than $\lambda\,$.

In our present work we investigate the accuracy of a mechanically driven single electron shuttle using a Master equation approach \cite{Lehrbuch}. The Master equation is solved numerically and also analytically in a simplified form. In particular, we calculate both the average number of electrons transferred per period which in general is different from the simple result (\ref{eq:N_average}) and the root mean square fluctuations $\Delta N \equiv\sqrt{\left<N^2\right>-\left<N\right>^2}\,$. As one of our main results, it turns out that optimum operation with $\Delta N\ll 1$ is obtained for long contact times ($t_0\gg RC$) and sufficiently low temperatures ($k_{\rm B}T\ll\frac{e^2}C$).

\section{Master equation}
In order to calculate the probability $p(m,t)$ to find $m$ additional electrons on the island at time $t\,$ we use a Master equation. This technique has been successfully used to describe both quantum dots and metallic islands exhibiting Coulomb blockade \cite{Gra91}-\cite{Bru94}. In our case the island oscillates mechanically which leads to time dependent transition rates $\Gamma(m,t)$ at the leads. Collecting gain and loss terms the Master equation reads
\begin{eqnarray}
  \frac{d}{dt} p(m,t) 
  =&-& 
  \left[\Gamma^{(+)}_{\rm L}(m,t)+\Gamma^{(+)}_{\rm R}(m,t)+\Gamma^{(-)}_{\rm L}(m,t)+
  \Gamma^{(-)}_{\rm R}(m,t)\right]p(m,t) \nonumber\nopagebreak\\
  &+&\left[\Gamma^{(+)}_{\rm L}(m-1,t)+\Gamma^{(+)}_{\rm R}(m-1,t)\right]p(m-1,t)\nonumber\nopagebreak\\
  &+&\left[\Gamma^{(-)}_{\rm L}(m+1,t)+\Gamma^{(-)}_{\rm R}(m+1,t)\right]p(m+1,t)\,.
\label{eq:MasterEq}
\end{eqnarray}
In a golden rule approach the tunnelling rates are of the form \cite{Lehrbuch}
\begin{equation}
 \Gamma = \frac 1 {e^2R} \, \frac {\Delta E}{1-\exp\left(-\frac{\Delta E}{k_{\rm B}T}\right)}
\,.
\end{equation}
Since both $R$ and the capacitance $C$ in $\Delta E \propto \frac{e^2}{C}$ are time dependent now, with the dominant time dependence associated with the exponentially varying $R(t)$, the tunnelling rates factorize into:
\begin{equation}
 \Gamma_{\rm R}^{(\mp)}(m,t) 
  = 
 g_{\rm R}(t)\,\Gamma_{{\rm R},m}^{(\mp)}(t) \,.
\end{equation}
Here
\begin{equation}
  g_{\rm R}(t) 
  =  
  \frac{R_{\rm R}(t_{\rm max})C_{\rm R}(t_{\rm max})}{R_{\rm R}(t)C_{\rm R}(t)}
\end{equation}
is a strongly varying function of time, dominated by the exponential decrease of tunnelling when the island moves away from the electrode ($t_{\rm max}$ is the time where the island is at its closest point to the right electrode). The motion is taken as purely harmonic, {\it i.e.\/} $x(t)=x_{\rm max}\sin(\omega t)$. For the rates $\Gamma_{{\rm R},m}^{(\mp)}(t)$ whose time dependence is determined by that of the capacitance only, we take the standard result for tunnelling rates when a voltage of $-V/2$ is applied at the left electrode and a voltage of $V/2$ is applied at the right electrode: 
\begin{equation}
\Gamma_{{\rm R},m}^{(\mp)}(t) = \frac 1{\tau} 
  \frac{\pm \left(m+ \frac{C_{\rm L}(t)V}{e}\right) - \frac 12}{1-\exp\left[-\left(\pm\left(m + \frac{C_{\rm L}(t)V}{e}\right)- \frac 12\right)\frac{e^2}{C_{\rm \Sigma}(t)k_{\rm B}T}\right]}
\end{equation}
where $\tau = R_{\rm R}(t_{\rm max})C_{\rm R}(t_{\rm max})$ and $C_{\rm \Sigma}(t)=C_{\rm R}(t)+C_{\rm L}(t)$. For the left electrode the indices $R$ and $L$ have to be interchanged, $V$ has to be replaced by $-V$ and $x(t)$ by $-x(t)\,$.  

For the numerical calculations the full Master equation was used. For simplicity, we assumed that the capacitance depends linearly on separation like $C_{\rm L}(t) \propto\frac 1{x_0+x_{\rm max}+x(t)}\,$. The function $g(t)$ is of the form
\begin{equation}
   g_{\rm R}(t) = \frac{x_0+x_{\rm max}-x(t)}{x_0}\exp\left[-\frac{x_{\rm max}-x(t)}{\lambda}\right]
\end{equation}
which is sharply peaked around $t_{\rm max}$.
Since tunnelling through the left and right electrodes does not take place simultaneously ($x_{\rm max}\gg\lambda$), the tunnelling at the left or the right electrodes can be treated separately. For the analytic calculations we consider only one electrode:
$
  \Gamma^{(\pm)}_{\rm L}(m,t)=0\,,
$
 which is approached by an island with $n$ electrons from the left, {\it i.e.\/} we are calculating the conditional probability $p(m|n,t)$ to find $m$ electrons on the island given that initially there where $n$ electrons on the island. 

It has to be noted that when the island is close to the right electrode, $C_{\rm L}(t)$ changes much slower than $C_{\rm R}(t)$ - independent of the precise model for the capacitance. As long as the applied voltage $V$ is not too large ({\it i.e.\/} as long as $VC_{\rm L}(t_{\rm max})$ is of order $e$, $2e$, \ldots) $C_{\rm L}(t)$ can be replaced by  $C_{\rm L}(t_{\rm max})\,$. In order to be able to solve the master equations analytically, we replace the function $g_{\rm R}(t)$ by a step function $\tilde{g}_{\rm R}(t)$ having the same height and area:
\begin{equation}
 \tilde{g}_{\rm R}(t) \equiv   
     \left\{
    \begin{array}{r@{\quad:\quad}l} 0&t\le t_{\rm max}-t_0\\    
         1 
 & t_{\rm max}-t_0<t<t_{\rm max}+t_0\\
 0 
 &t \ge t_0+t_{\rm max} 
    \end{array}
    \right.\,,\hspace{1cm}
 t_0 \equiv \sqrt{\frac{\pi\lambda}{2x_{\rm max}}}\frac 1{\omega}\left(1+\frac {\lambda}{2x_0} \right)\,.
\label{eq:gres}
\end{equation}
 The width $2t_0$ corresponds to an effective contact time.
 As expected, the contact time decreases for increasing oscillation frequencies ($x_{\rm max} \gg \lambda\,$, $t_{\rm max}\gg t_0$).

The solutions $p(m|n,t)$ of the Master equation can be used to calculate the probability $p_{\Delta}(m)$ that $m$ electrons are transferred:
\begin{equation}
  p_{\Delta}(n) 
  \equiv \sum_{k} 
   p_{\rm i}(k)p({k-n|k},t_0+t_{\rm max})\,,
\label{eq:p_D_def}
\end{equation}
where $p_{\rm i}(k)$ is the initial probability to find $k$ additional electrons on the island ($\sum_{k}p_{\rm i}(k)=1$). 
To calculate this probability, we consider a symmetric situation where the probability for the island to approach the right electrode with $n$ electrons is the same as the probability for the island to approach the left electrode with $-n$ electrons:
\begin{equation}
  p_{\rm i}(n) = p_{\rm f}(-n)\,,
\end{equation}
 where
\begin{equation}
 p_{\rm f}({m})=\sum_{n} p_{\rm i}({n})p({m|n},t_0+t_{\rm max})\,.
\label{eq:p_f_eq}
\end{equation}

This leads to a set of linear equations for $p_{\rm i}({n})$ which in principle can  be solved for arbitrary values of $n_{\rm max}$ and $t_0$. However, for practical purposes, small values for $n_{\rm max}$ should be used. From $p_{\Delta}(n)$ we then calculate both $\left< N\right>$ and $\Delta N$ ($\left< N^k\right>\equiv \sum_NN^kp_{\Delta}(N)$).

\section{Low temperature limit}
For very low temperatures $k_{\rm B}T\ll\frac{e^2}{C}$ Coulomb blockade fixes the number of additional electrons $n$ on the island between $-n_{\rm max}$ and $n_{\rm max}$ strictly. We will focus on the behaviour of the single electron shuttle for small integer values of $n_{\rm max}$ ($\frac{C_{\rm L}(t_{\rm max})V}{e} = n_{\rm max}+\delta$; $-\frac 12\le\delta< \frac 12$).

With the approximation described in eq.~(\ref{eq:gres}) the Master equation at $T=0$ simplifies to :
\begin{equation}
\frac{d}{dt}{p}({n-k}|n,t) = -\frac 1{\tau_{n-k}}{p}({n-k}|n,t)
+\frac 1{\tau_{n-k+1}}{p}(n-k+1|n,t)
\label{eq:master_tilde}
\end{equation}
with 
\begin{equation}
  \frac{1}{\tau_{-n_{\rm max}+\nu}}
  =
\left\{
    \begin{array}{r@{\quad:\quad}l}\frac 1{\tau}\left(\nu - \frac 12+\delta\right) &\nu>0\\
 0 
 & \nu\le0
    \end{array}\right.\,.
\end{equation}
We solve eq.~(\ref{eq:master_tilde}) for $t_{\rm max}-t_0<t<t_{\rm max}+t_0\,$:
\begin{equation}
  {p}(n-k|n,t) 
  = 
  \frac 1{k!}\left(\prod_{i=n-k+1}^n \frac{\tau}{\tau_{\rm i}}\right)
  \exp\left(-\frac{\Delta t}{\tau_{n-k}}\right)
  \left[1-\exp\left(-\frac{\Delta t}{\tau}\right)\right]^k
\label{eq:p_T0}
\end{equation}
where $\Delta t\equiv t_{\rm max}-t_0+t$ ($n, n-k>-n_{\rm max}$). Eq.~(\ref{eq:p_T0}) can easily be proved by induction using the fact that $\frac 1{\tau_m}+\frac 1{\tau}=\frac 1{\tau_{m+1}}$. For $n=n_{\rm max}\,$, the normalization condition leads to the following result: ${p}(-n_{\rm max}|n,t)=1-\sum_{m=-n_{\rm max}+1}^n p(m|n,t)\,$. 

\begin{figure}
\vbox to 5.25cm {\vss\hbox to 3.0cm
 {\hss\
   {\includegraphics{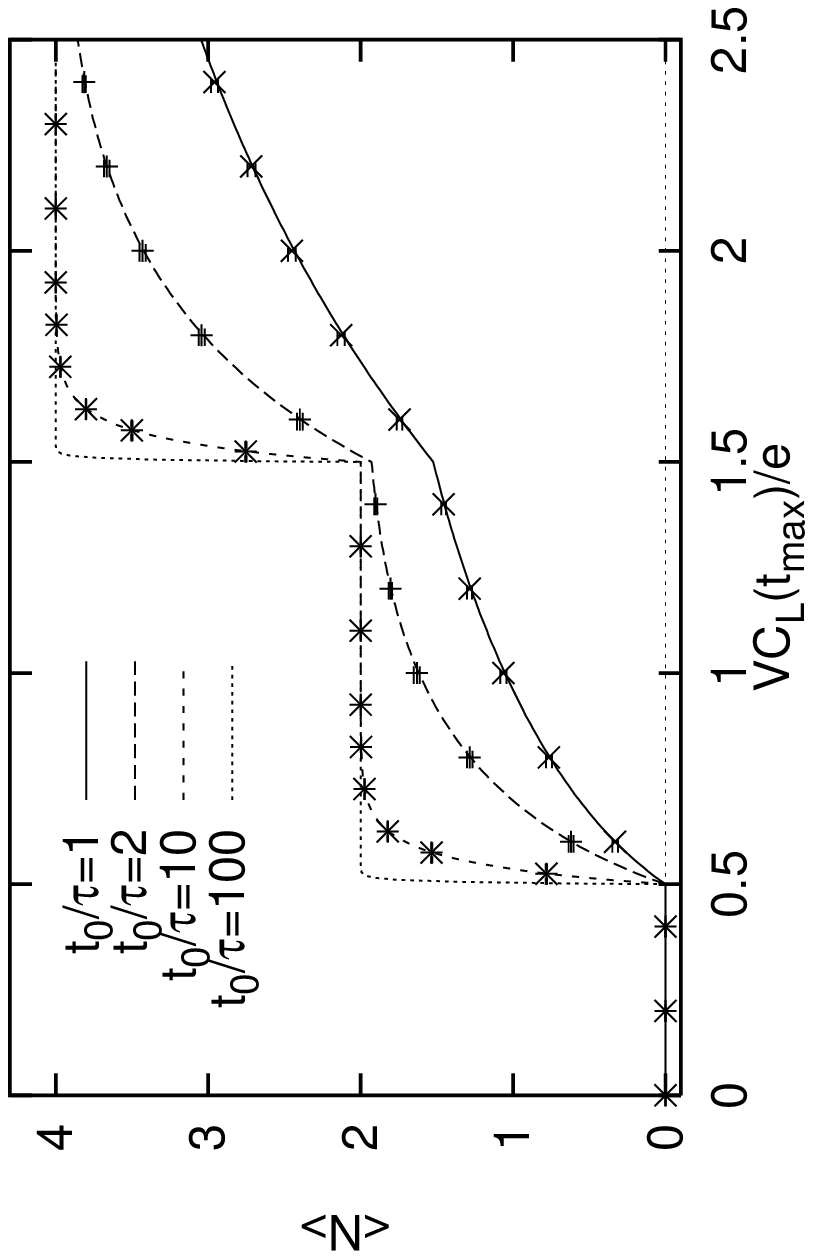}
  }
   {\includegraphics{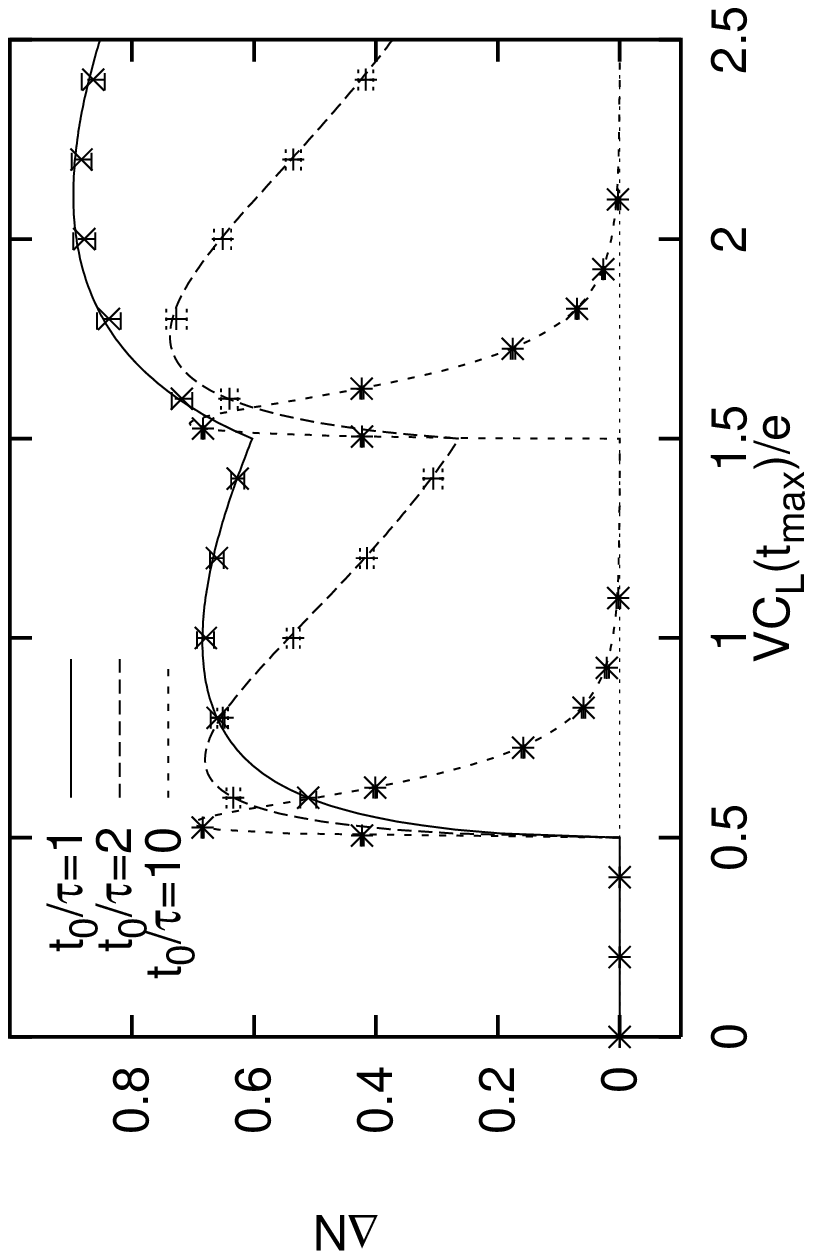}
   }
    \hss}
 }
\caption{The average number of electrons transferred per period (left) and the root mean fluctuations (right) for $T=0$. Coulomb blockade is clearly visible: up to a critical voltage ($\frac{VC_{\rm L}(t_{\rm max})}{e} = \frac 12$) no electrons are transferred. The Coulomb staircase becomes symmetric for $t_0\gg\tau\,$. The agreement between analytical results (lines) and computer simulations (crosses) is excellent.}
\label{fig:coulombblockade}
\end{figure}

\begin{figure}
\vbox to 5.0cm {\vss\hbox to 3.0cm
 {\hss\
   {\includegraphics{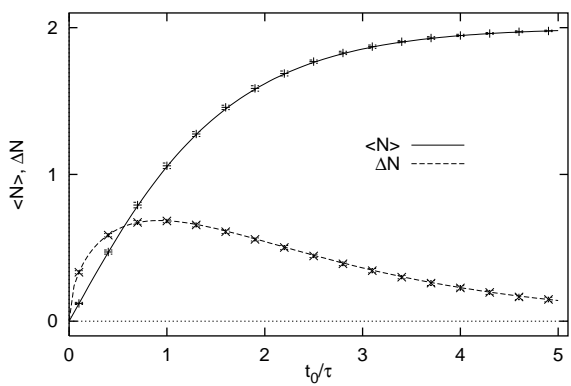}
  }
    \hss}
 }
\caption{The average number of electrons transferred per period (solid line) and the root mean fluctuations (dashed line) for $T=0$ and $\frac{VC_{\rm L}(t_{\rm max})}{e} = 1$. For $t_0\ll t_{\rm max}$ no current flows, the optimum range is for $t_0\gg t_{\rm max}$ where the fluctuations disappear.}
\label{fig:NT0}
\end{figure}

Together with eqs.~(\ref{eq:p_D_def}) and (\ref{eq:p_f_eq}) this can be used to calculate the current and the current fluctuations for $T=0\,$. The results obtained from the analytical expression~(\ref{eq:p_T0}) and also from the numerical solution of the full master equation are shown in fig.~\ref{fig:coulombblockade}. Coulomb blockade is clearly visible: for low voltages no electrons are transferred. Above a critical voltage of $V=\frac {e^2}{2C_{\rm L}(t_{\rm max})}$ the shuttle starts to work.
For $n_{\rm max}=1$, $\delta = 0$ the analytic expressions for the average number of electrons transferred ($\left<N\right>$) and the mean square fluctuations ($(\Delta N)^2$) are comparatively simple: 
\begin{eqnarray}
  \left<N\right> &=& 
  2\frac{1-a^3}
  {\left[1+a\right]\left[1+\frac 12a+a^2\right]}
\label{eq:NmeanT0}
\\
 (\Delta N)^2 &=& 2a(1-a)\frac{6+9a+22a^2+13a^3+10a^4}{\left[2a^2+a+2\right]^2\left[a+1\right]^2}
\end{eqnarray}
where 
$
 a\equiv \exp\left(-\frac{t_0}{\tau}\right)\,.
$
As expected, in the limit of very short contact times $t_0\ll \tau$ ($a\rightarrow 1$), no electrons are transferred whereas very long contact times $t_0\gg\tau$ ($a\rightarrow 0$) lead to the simple result~(\ref{eq:N_average}). The fluctuations disappear for both short and long contact times. Their maximum ($\Delta N\approx 0.686 $) is reached for $t_0\approx 0.92 \tau$ as shown in fig.~\ref{fig:NT0}.

\section{Finite temperatures}
For finite temperatures the number of additional electrons on the island is no longer as restricted as for $T=0$. Here, the Master equation has the more complicated form:
\begin{eqnarray}
 \frac{\rm d}{{\rm d}t}p(-n_{\rm max}\!+\!k|n,t)
 =  
 \frac {g(t)}{\tau}\frac{k-\frac {1}{2}+\delta}{\left(1-\varepsilon^{2k-1}\xi\right)}
   \left[\varepsilon^{2k-1}\xi p(-n_{\rm max}\!+\!k\!-\!1|n,t) -p\left(-n_{\rm max}\!+\!k|n,t\right)\right]
\nonumber\\
 +  \frac {g(t)}{\tau} \frac{k+\frac {1}{2}+\delta}{\left(1-\varepsilon^{2k+1}\xi\right)} \left[p(-n_{\rm max}\!+\!k\!+\!1|n,t) -{\varepsilon^{2k+1}}\xi p(-n_{\rm max}\!+\!k|n,t)\right]\,,
\end{eqnarray}
where $\varepsilon \equiv \exp\left(-\frac{e^2}{2C_{\rm \Sigma}(t)k_{\rm B}T}\right)$ and  $\xi = \varepsilon ^{2\delta}\,$.
For $t_0\gg \tau$ and constant $C_{\rm \Sigma}$ ({\it i.e.\/} constant $\varepsilon$ and $\xi$) the stationary solution ($\frac{d}{dt}p(m|n,t)=0$, which is independent of $n$ ($p_{\rm f}(m)=p(m|n,t_{\rm max}+t_0)$) is given by:
\begin{equation}
  p_{\rm f}(-n_{\rm max}+k)
  = \eta^{-1}\,{\varepsilon^{k^2}}\xi^k
\,,
\label{eq:p_fT}
\end{equation}
 with the normalization constant 
$ {\eta\equiv \sum_{\nu=-\infty}^{\infty}\varepsilon^{\nu^2}\xi^{\nu}}\,.$ Together with eq.~(\ref{eq:p_D_def}) this leads to:
\begin{equation}
 p_{\Delta}(2n_{\rm max}+k) =\frac{\xi^{-k}}{\eta^2} {\sum_{\nu=-\infty}^{\infty}\varepsilon^{\nu^2}\varepsilon^{(\nu+k)^2}}{}
 \,. 
\label{eq:p_deltaT}
\end{equation}
In reality however, $C_{\rm \Sigma}(t)$ is time-dependent and thus the results (\ref{eq:p_fT}) and (\ref{eq:p_deltaT}) are not exact. It turns out however, that if one replaces $C_{\rm \Sigma}(t)$ by $C_{\rm \Sigma}(t^*)$ where $t^*$ is defined by
\begin{equation}
  \int_{t^*}^{\pi/\omega}g(t){\rm d}t = \tau
\label{eq:def_t}
\end{equation}
one obtains good agreement between numerical results obtained by the solution of the time dependent problem and the analytic result (\ref{eq:p_fT}). (Changes which happen on time-scales much smaller than $\frac{\pi}{\omega}-t^*$, {\it i.e.\/} effective time-scales much smaller than $\tau$, are too fast to be followed, cf.\ fig.~\ref{fig:NT0}.)

\begin{figure}
\vbox to 5.25cm {\vss\hbox to 3.0cm
 {\hss\
   {\includegraphics{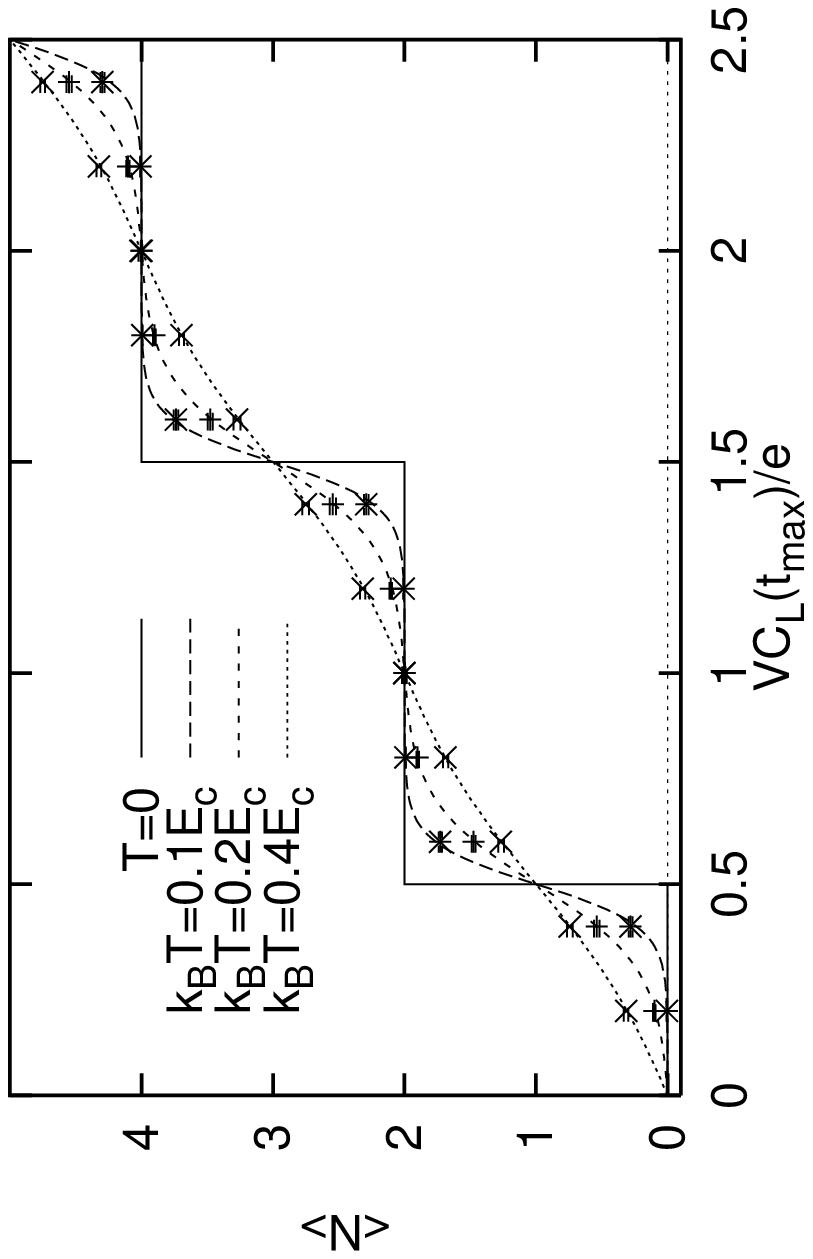}
  }
   {\includegraphics{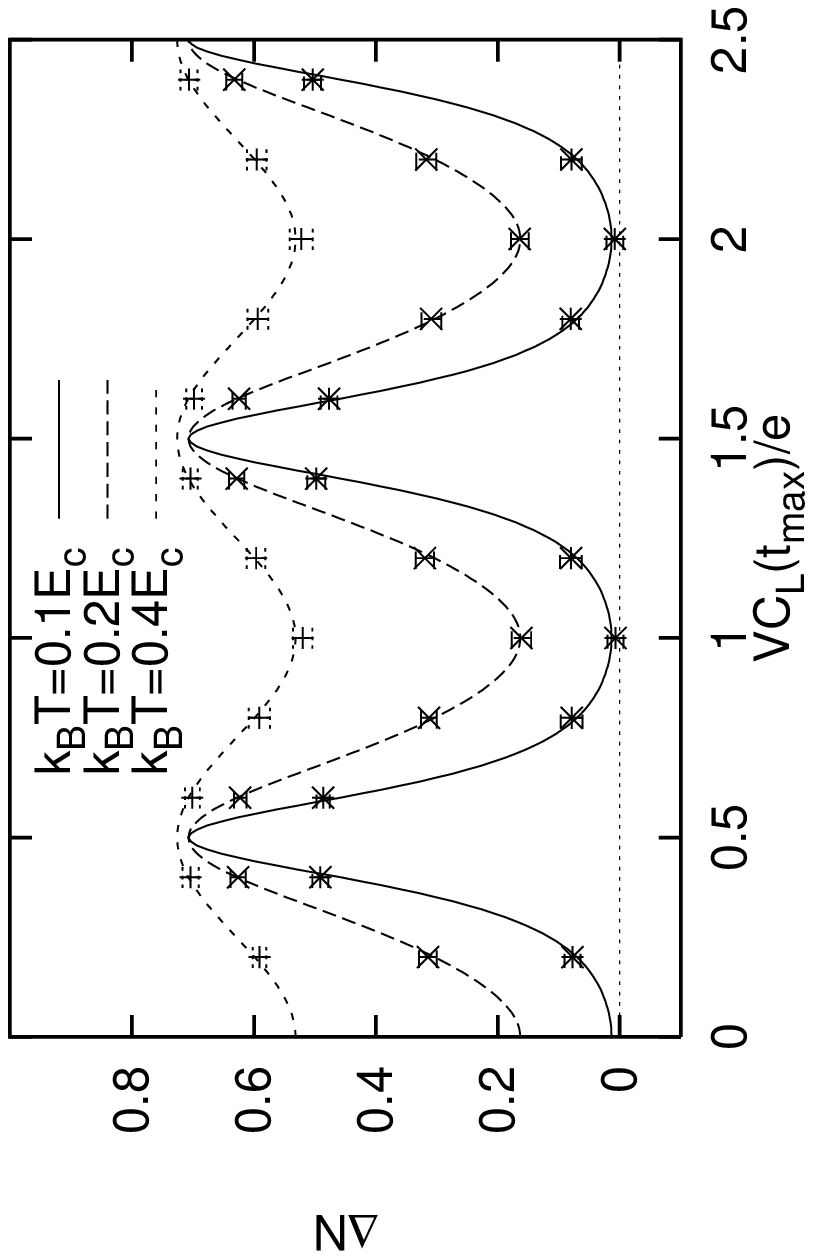}
   }
    \hss}
 }

\caption{For $t_0\gg\tau$ and $T=0$ (solid line) the Coulomb staircase is perfect. For increasing temperatures, the Coulomb-staircase is still symmetric, however it is washed out leading to an ohmic behaviour for high temperatures. (The energy scale is given by $E_c=\frac{e^2}{2C_{\rm \Sigma}}$)}
\label{fig:coulombTg0}
\end{figure}

Fig. \ref{fig:coulombTg0} shows two steps of the symmetric resulting Coulomb staircase for several temperatures. In the middle of one of the steps of the Coulomb staircase ($\delta =0$) the average number of electrons transferred per period agrees with the simple result 
$
 \left< N \right> 
 = 
 2n_{\rm max}
$
even for $k_{\rm B}T>\frac {e^2}{2C_{\rm \Sigma}}\,$.
The root mean square fluctuations are approximately given by 
\begin{equation}
 \Delta N
 \simeq \left\{
    \begin{array}{r@{\quad:\quad}l}   
         2\sqrt{\frac{\varepsilon}{2\varepsilon+1}}
 & k_{\rm B}T\ll\frac{e^2}{2C_{\rm \Sigma}}\\
         \sqrt{k_{\rm B}T/\frac{e^2}{2C_{\rm \Sigma}}}
 & k_{\rm B}T\gg \frac{e^2}{2C_{\rm \Sigma}}
    \end{array}
    \right.\,,
\hspace{1cm}\varepsilon 
   =
 \exp\left(-\frac{e^2}{2C_{\rm \Sigma}(t^*)k_{\rm B}T}\right)\,.
\end{equation}
As expected, the fluctuations disappear exponentially for decreasing temperatures.

\section{Summary}
Using a simple Master equation we have derived analytic expressions for both the average current and its root mean square fluctuations in a mechanically driven single electron shuttle.  We find a  good agreement between our analytical results and a numerical solution of the fully time-dependent Master equation.

To minimize fluctuations the shuttle should be operated under conditions where the interaction time $t_0$ is much greater than the $RC$-time  $\tau$ ($t_0\approx 10 \tau)$) and the temperature $k_{\rm B}T$ is much lower than the Coulomb energy $\frac {e^2}{2C_{\rm \Sigma}}$.

We would like to thank R.~Blick, A.~Erbe and N.~Chandra for useful discussions.
Support by the {\it  SFB 348 }of the {\it Deutsche Forschungsgemeinschaft }is greatfully acknowledged.

\newpage

\end{document}